\documentclass[letterpaper,10pt,oneside,conference,final]{IEEEtran}
\usepackage[latin1]{inputenc}
\usepackage[pdftex]{graphicx}
\usepackage{color}
\usepackage{booktabs}
\usepackage{subfig}
\usepackage{tikz,pgfplots}
\usetikzlibrary{arrows.meta}
\usetikzlibrary{spy}
\usepackage{amssymb}
\usepackage{amsmath}
\usepackage{tabularx}
\usepackage{makecell}
\graphicspath{{figures/}}
\def\BibTeX{{\rm B\kern-.05em{\sc i\kern-.025em b}\kern-.08em
		T\kern-.1667em\lower.7ex\hbox{E}\kern-.125emX}}

\begin{document}
	\IEEEoverridecommandlockouts
	\title{In-phase and Quadrature Chirp Spread Spectrum for IoT Communications}
	
	\author{\IEEEauthorblockN{Ivo Bizon Franco de Almeida\IEEEauthorrefmark{1}, Marwa Chafii\IEEEauthorrefmark{2}, Ahmad Nimr\IEEEauthorrefmark{1} and Gerhard Fettweis\IEEEauthorrefmark{1}}
		\IEEEauthorblockA{\IEEEauthorrefmark{1}Vodafone Chair Mobile Communications Systems, Technische Universit{\"a}t Dresden (TUD), Germany \\
			Email: \{ivo.bizon, ahmad.nimr, gerhard.fettweis\}@ifn.et.tu-dresden.de}
		\IEEEauthorblockA{\IEEEauthorrefmark{2}ETIS, UMR8051, CY Cergy Paris Université, ENSEA, CNRS, France \\
			Email: marwa.chafii@ensea.fr}
	}
	
	\maketitle
	
	\begin{abstract}
		~ This paper describes a coherent chirp spread spectrum (CSS) technique based on the Long-Range (LoRa) physical layer (PHY) framework. 
		LoRa PHY employs CSS on top of a variant of frequency shift keying (FSK), and non-coherent detection is employed at the receiver for obtaining the transmitted data symbols. 
		In this paper, we propose a scheme that encodes information bits on both in-phase and quadrature components of the chirp signal, and rather employs a coherent detector at the receiver. 
		Hence, channel equalization is required for compensating the channel induced phase rotation on the transmit signal. 
		Moreover, a simple channel estimation technique exploits the LoRa reference sequences used for synchronization to obtain the complex channel coefficient used in the equalizer. 
		Performance evaluation using numerical simulation shows that the proposed scheme achieves approximately 1 dB gain in terms of energy efficiency, and it doubles the spectral efficiency when compared to the conventional LoRa PHY  scheme. 
		This is due to the fact that the coherent receiver is able to exploit the orthogonality between in-phase and quadrature components of the transmit signal.
	\end{abstract}
	
	
	\section{Introduction}
	
	\IEEEPARstart{R}{ecently}, a lot of attention has been drawn towards long-range and low-power consuming wireless communication schemes \cite{lpwans}. 
	Long-Range (LoRa) is a wireless communication protocol that has got preference among the schemes considered primary for Internet of things (IoT) applications \cite{survey1}. 
	The main application of LoRa, and low power wide area networks (LPWAN) in general, is to provide connectivity for mobile and stationary wireless end-devices that require data rates in the order of tens of kbps up to a few Mbps within a coverage area up to tens of kilometers. 
	Low energy consumption and simple design are also desirable characteristics for LPWAN devices \cite{iot_citation}.
	
	The physical layer (PHY) of LoRa has gained considerable attention of the academic community, and several papers have been published with investigations on the characteristics of LoRa PHY and MAC schemes \cite{lora_limits,lora_capacity,lora_sf_orthogonality}.
	
	Some authors have proposed enhancements to the LoRa PHY framework.
	For instance, encoding extra information bits on the phase of the chirp waveform has been proposed in \cite{psk_lora}, and similar work makes use of pulse shaping the chirp waveform to reduce the guard-band, and thus increases the number of channels within the available frequency band \cite{efficient_css}.
	
	Inspired by the aforementioned works, and departing from the LoRa PHY framework, we propose to extend it to a coherent scheme that is able to increase the spectral efficiency (SE) (bps/Hz) and improve energy efficiency (EE).
	We define this scheme as in-phase and quadrature chirp spread spectrum (IQCSS), since information is encoded in both in-phase (real) and quadrature (imaginary) components of the transmit signal.
	
	The remainder of the paper is organized as follows: 
	Section II presents a background on chirp spread spectrum (CSS).
	Section III describes the proposed IQCSS scheme together with some implementation remarks.
	Section IV presents the performance evaluation of the proposed scheme in terms of the bit error ratio.
	Finally, section V concludes the paper.
	
	\section{Background Chirp Spread Spectrum}
	
	\begin{figure*}[ht!]
		\centering
		\includegraphics[]{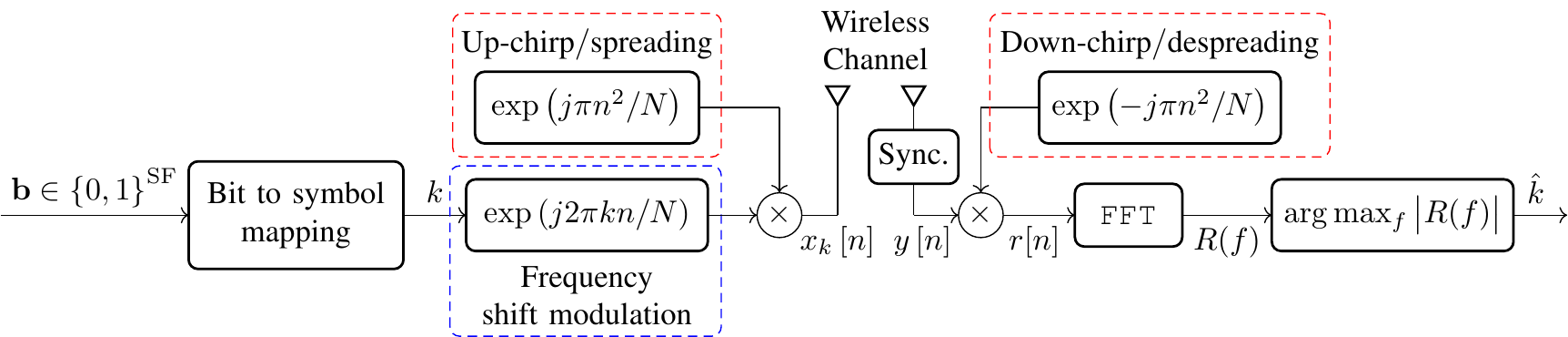}
		\caption{LoRa PHY transceiver block diagram.}
		\label{LoRATXRX}
	\end{figure*}
	
	This section aims to give an analytical description of the chirp signal as well as its implementation on LoRa PHY.
	
	\subsection{The chirp signal}
	In CSS \cite{css1}, as well as in other spread spectrum techniques, such as direct sequence (DSSS) and frequency hoping (FHSS), the information is transmitted using a bandwidth much larger than required for a given data rate. 
	Particularly in CSS, multiplication by a chirp signal is responsible for the energy spreading in frequency. 
	The linear-chirp refers to the frequency variation of the signal, which increases linearly with time. 
	
	The chirp waveform employed in CSS can be described by
	\begin{equation}
	c(t) = \begin{cases}
	\exp\left( j\varphi(t) \right) & \text{for } -T/2 \leq t \leq T/2 \\
	0 & \text{otherwise,}
	\end{cases}
	\end{equation}
	where $\varphi(t) = \pi (at^2 + 2bt)$, i.e., a quadratic function of time. 
	The chirp instantaneous frequency is defined as 
	\begin{equation}
	v(t) = \frac{1}{2\pi} \frac{d \varphi(t)}{dt} = at + b,
	\end{equation}
	which shows that the frequency varies linearly with time. 
	Moreover, the chirp rate is defined as the second derivative of $\varphi(t)$ w.r.t. $t$ as
	\begin{equation}
	u(t) = \frac{1}{2\pi} \frac{d^2 \varphi(t)}{dt^2} = a.
	\end{equation}
	The up-chirp is defined when $u(t)>0$, and the down-chirp when $u(t)<0$.
	
	Let $B$ (Hz) represent the bandwidth occupied by the chirp signal.
	The signal frequency varies linearly between $-B/2 + b$ and $B/2 + b$ within the time duration $T$ (s).
	If the term $b=0$ and $a \neq 0$, the resulting waveform is the \textit{raw} chirp with starting frequency $-B/2$ and end frequency $B/2$. 
	Conversely, if $a=0$, the complex exponential is obtained.
	After sampling at rate $B = 1/T_s$, where $T_s$ (s) is sampling time interval, the discrete-time chirp signal is given by
	\begin{equation}
	c(nT_s) = \begin{cases}
	\exp\left( j\varphi(nT_s) \right) & \text{for } n = 0, \ldots, \, N-1 \\
	0 & \text{otherwise,}
	\end{cases}
	\end{equation}
	where $N = T/T_s$ is the total number of samples within $T$. Setting $b=0$, and $a = B/T$, the discrete-time raw up-chirp becomes
	\begin{equation}
	c[n] = \exp\left( j\pi n^2/N \right).
	\end{equation}
	
	\subsection{LoRa PHY}
	
	Notably, LoRa PHY employs CSS in conjunction with a variant of frequency shift keying (FSK) modulation \cite{the_lora_modulation,loradatasheet}. 
	At the transmitter side, $c[n]$ is used for spreading the information signal within the bandwidth $B$ via multiplication.
	The despreading operation at the receiver side is accomplished by multiplying the received signal with a down-chirp, which is obtained by conjugating the up-chirp.
	The transmit signal can be described as 
	\begin{equation}
	x_k[n] = \sqrt{\frac{E_s}{N}} \exp \left( j\frac{2\pi }{N}kn \right) c[n],
	\end{equation}
	where the exponential term has its frequency depending upon the data symbol $k$.
	LoRa PHY defines the spreading factor (SF) as the amount of bits that one symbol carriers, which ranges from 6 to 12 bits. 
	Note that each waveform has $N = 2^{\mathrm{SF}}$ samples for having distinguishable waveforms. 
	The data symbols are integer values from the set $ \mathbb{K} = \left\lbrace 0, \ldots, \, 2^{\mathrm{SF}} - 1\right\rbrace $, which contains $N$ elements. 
	$E_s$ represents the signal energy. 
	
	The spreading gain, also known as processing gain, is defined by the ratio between the bandwidth of the spreading chirp signal and the information signal, and it can be defined in dB as 
	\begin{equation}
	G = 10\log_{10}\left(\frac{N}{\mathrm{SF}}\right).
	\end{equation}
	
	In short, Fig. \ref{LoRATXRX} shows the LoRa transceiver block diagram. 
	The code-word $\mathbf{b}$ contains SF bits that are mapped into one symbol $k$, which feeds the CSS modulator. 
	At the receiver side, the estimated data symbol is obtained by selecting the frequency index with maximum value. 
	This operation can be described as
	\begin{equation}
	\hat{k} = \arg\max_{f \in \mathbb{K}} \big| R(f)  \big| ,
	\end{equation}
	where $R(f) = \mathcal{F}\left\lbrace r[n]\right\rbrace$, $r[n]$ represents the received signal after despreading, and $\mathcal{F}\left\lbrace \cdot \right\rbrace$ the discrete Fourier transform. 
	Luckily, this operation can be easily carried via the fast Fourier transform (FFT) algorithm.
	
	A key aspect of LoRa PHY is the fact that channel estimation and equalization are not necessary, since it employs a non-coherent FSK. 
	However, employing coherent detection can improve the performance of LoRa.
	
	\section{In-phase and Quadrature Chirp Spread Spectrum}
	
	\begin{figure*}[ht!]
		\centering
		\includegraphics[]{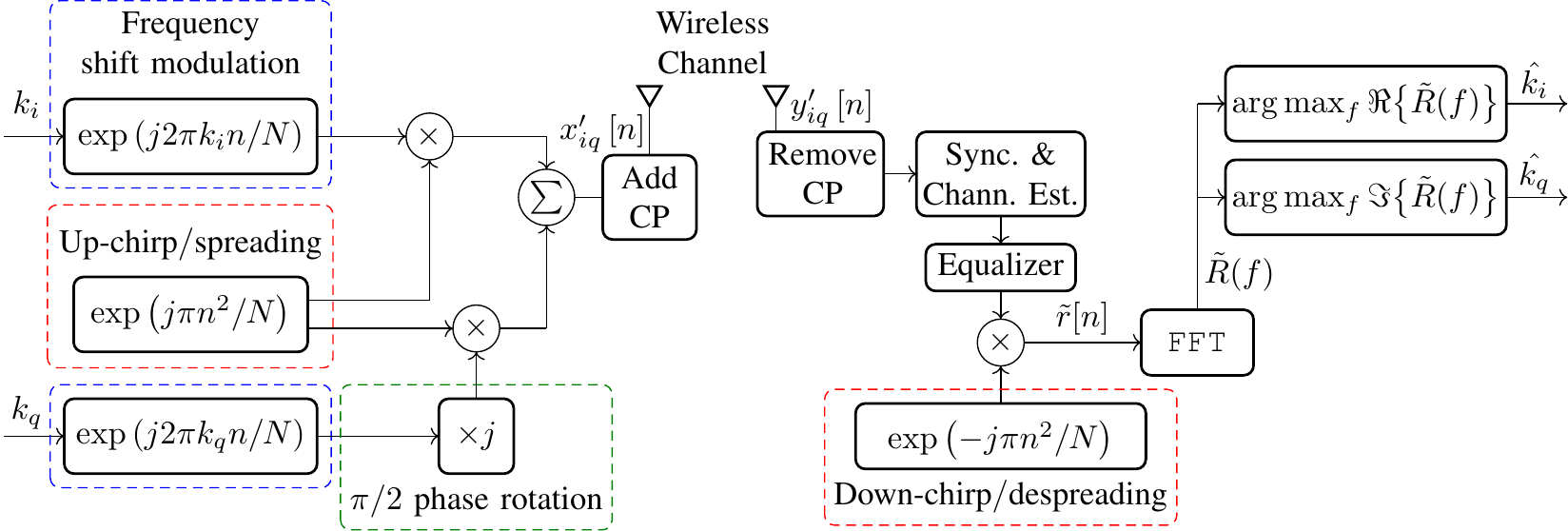}
		\caption{IQCSS transceiver block diagram.}
		\label{IQCSS_TXRX}
	\end{figure*}
	
	The synchronization preamble used in LoRa's PHY makes use of 8 up-chirps for synchronization, followed by 2 down-chirps to indicate the start of the data symbols \cite{loradatasheet}. 
	To make further use of the information carried with the synchronization signal, we propose to use the least squares (LS) approach for estimating the channel gain using the already available structure for synchronization \cite{kay1993fundamentals}. 
	The received preamble can be modeled as
	\begin{equation}
	y_p[n] = hx_p[n] + w[n],
	\end{equation}
	where $x_p[n]$ represents the 8 raw up-chirps transmitted, $w[n]$ is additive white Gaussian noise (AWGN) with zero mean and $\sigma^2_w$ variance.
	Assuming that the channel presents flat-fading within its bandwidth, the LS estimator of the complex channel gain is given by
	\begin{equation} \label{channel_estimator}
	\hat{h} = \frac{\mathbf{x}_p^{\mathrm{H}}\mathbf{y}_p}{\mathbf{x}_p^{\mathrm{H}}\mathbf{x}_p},
	\end{equation}
	where $\mathbf{x}_p$ and $\mathbf{y}_p$ are $N_p\!\times\!1$ vectors whose entries are the samples from $x_p[n]$ and $y_p[n]$, respectively. $\left(\cdot\right)^{\mathrm{H}}$ represents the Hermitian operation, $N_p=8N$ is the length of the preamble in samples, and $h$ and $\hat{h}$ are the true and estimated complex channel gain, respectively. 
	
	For the case of frequency selective channels the LS approach can be extended, but in this case a cyclic prefix (CP) needs to be added to the synchronization chirps. 
	Under the assumption of a CP, the estimation of the equivalent channel impulse response can be obtained in vectorized form as
	\begin{align} \label{channel_estimator_selective_inv}
	\hat{\mathbf{h}} = \left( \mathbf{C}^{\mathrm{H}} \mathbf{C} \right)^{-1}\mathbf{C}^{\mathrm{H}} \mathbf{\bar{y}}_p,
	\end{align}
	where $\mathbf{C}$ is a $N\!\times\!N$ circulant matrix obtained from one raw up-chirp, $\mathbf{\bar{y}}_p$ is a $N\!\times\!1$ vector whose entries are the averaged samples from the 8 received synchronization chirps after CP removal, and $\hat{\mathbf{h}}$ contains the estimated channel impulse response. 
	By inspecting (\ref{channel_estimator_selective_inv}) one can see that $\mathbf{C}$ is an orthogonal matrix, and the estimator simplifies as 
	\begin{align} \label{channel_estimator_selective}
	\hat{\mathbf{h}} = \mathbf{C}^{\mathrm{H}} \mathbf{\bar{y}}_p,
	\end{align}
	which reduces considerably the estimation computational complexity. 
	
	On the condition that the channel estimation and equalization modules are available at the receiver, coherent detection of CSS modulation can be made possible.
	In this case, the receiver structure can be modified for achieving greater SE and EE when compared with LoRa modulation. 
	
\begin{figure}[ht!]
		\centering
		\includegraphics[]{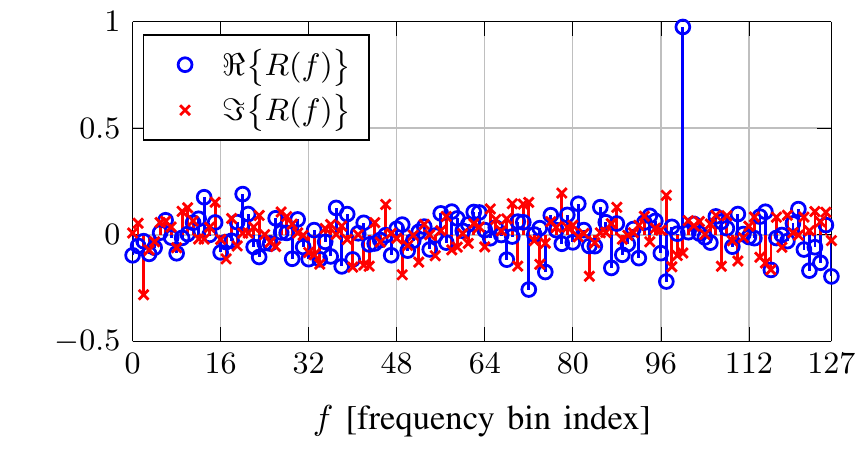}
		\caption{Real and imaginary parts of $R(f)$ considering $\mathrm{SF} = 7$, $k = 100$, and $\mathrm{SNR} = 10$ dB.}
		\label{RE_IM_RX_INDEX}
	\end{figure}

	For demonstrating how the EE can be improvement, let us consider the signal after the FFT in Fig. \ref{LoRATXRX}, which is given by
	$R(f) = \mathcal{F}\left\lbrace r[n]\right\rbrace$. 
	Figure \ref{RE_IM_RX_INDEX} shows the real and imaginary parts of $R(f)$.
	As one  can see, the information lies in the real part, and taking absolute value of $R(f)$ means collecting extra noise from the imaginary part. 
	Therefore, if the phase rotation induced by the channel is corrected, i.e., equalization is done before demodulation, the transmitted data symbol can be estimated as
	\begin{equation}
	\hat{k} = \arg\max_{f \in \mathbb{K}}  \Re\big\lbrace R(f)\big\rbrace,
	\end{equation}
	which is obtained with maximum likelihood estimation.
	
	Furthermore, the SE can be doubled if more information is also encoded in the quadrature component. 
	Taking this points into consideration, we define this scheme as IQCSS, and its transmit signal is given by
	\begin{equation} \label{tx_signal_IQCSS}
	x_{iq}[n] = \sqrt{\frac{E_s}{2N}} g_{iq}[n] c[n],
	\end{equation}
	where 
	\begin{equation} \label{index_signal_IQCSS}
	g_{iq}[n] = \exp \left( j\frac{2\pi }{N}k_in \right) + j\exp \left( j\frac{2\pi }{N}k_qn \right),
	\end{equation}
	where $k_i$ and $k_q$ are independent identically (uniform) distributed data symbols drawn from $\mathbb{K}$, and each carries $\mathrm{SF}$ bits.
	Here we make use of the orthogonality between the sine and cosine waves to transmit simultaneously two data symbols, and hence doubling the SE when compared to LoRa PHY. 
	Another key aspect of this approach is that receiving devices (gateways) that operate using IQCSS can still decode the information transmitted using the original LoRa modulation, thus making IQCSS gateways backwards compatible. 
	These characteristics are certainly attractive from the market perspective. 
	
	Figure \ref{IQCSS_TXRX} illustrates the proposed IQCSS transceiver block diagram. 
	Note that the additional operations do not require modifications on the data packet structure, since IQCSS makes use of the already available synchronization preamble for channel estimation. 
	This holds true under assumption of flat-fading channels, and channel equalization is performed with a simple complex multiplication.
	However, for frequency selective channels the addition of a CP becomes necessary, and in this case, low-complexity frequency domain equalization can be employed.
	
	The received signal after equalization and despreading is given by
	\begin{equation}
	\tilde{r}[n] = g_{iq}[n] + \tilde{w}[n],
	\end{equation}
	and the estimated data symbols are given by
	\begin{equation}
	\hat{k_i} = \arg\max_{f \in \mathbb{K}} \Re\big\lbrace \tilde{R}(f)\big\rbrace,
	\end{equation}
	\begin{equation}
	\hat{k_q} = \arg\max_{f \in \mathbb{K}} \Im\big\lbrace \tilde{R}(f)\big\rbrace,
	\end{equation}
	where $\tilde{R}(f)$ represents the discrete Fourier transform of $\tilde{r}[n]$.
	
	\section{Performance Evaluation}
	
	In order to characterize the performance of the proposed scheme, we resort to numerical simulations for estimating the bit error ratio (BER) under three channel models, namely AWGN, time-variant (TV) non-frequency-selective (Rayleigh) channel, and time-variant frequency-selective (TVFS). For the latter, the channel model defined as "Typical case for urban area" with 12 taps is employed \cite{gsm_channel}. This has been chosen due to the similar operating frequency of LoRa and Global System for Mobile Communications (GSM).
	Moreover, based on the error performance achieved under AWGN channel, the improvement with respect to spectral efficiency and increased throughput is also shown in this section.
	
	Table \ref{simulation_parameters} presents the simulation parameters considered for this section.
	\begin{table}[h]
		\centering
		\caption{Simulation parameters}
		\label{simulation_parameters}
		\renewcommand{\arraystretch}{1.5}
		\begin{tabularx}{7.85cm}{ll}
			\toprule[0.9pt] 
			Parameter & Value \\ 
			\midrule
			Spreading factor & $\mathrm{SF} \in \left\lbrace 6, \; \cdots, \; 12\right\rbrace$ \\
			Bandwidth & 250 kHz \\ 
			Carrier frequency & 863 MHz \\ 	
			Mobile speed & 0.1 km/h \\	
			CP length & 16 samples \\ 
			TV channel & single tap (Rayleigh) \\
			TVFS channel & 12 taps "Typical case for urban area" \cite{gsm_channel} \\
			\bottomrule[0.9pt]	
		\end{tabularx}
	\end{table}
	
	\subsection{AWGN channel}
	\begin{figure}[ht!]
		\centering
		\includegraphics[]{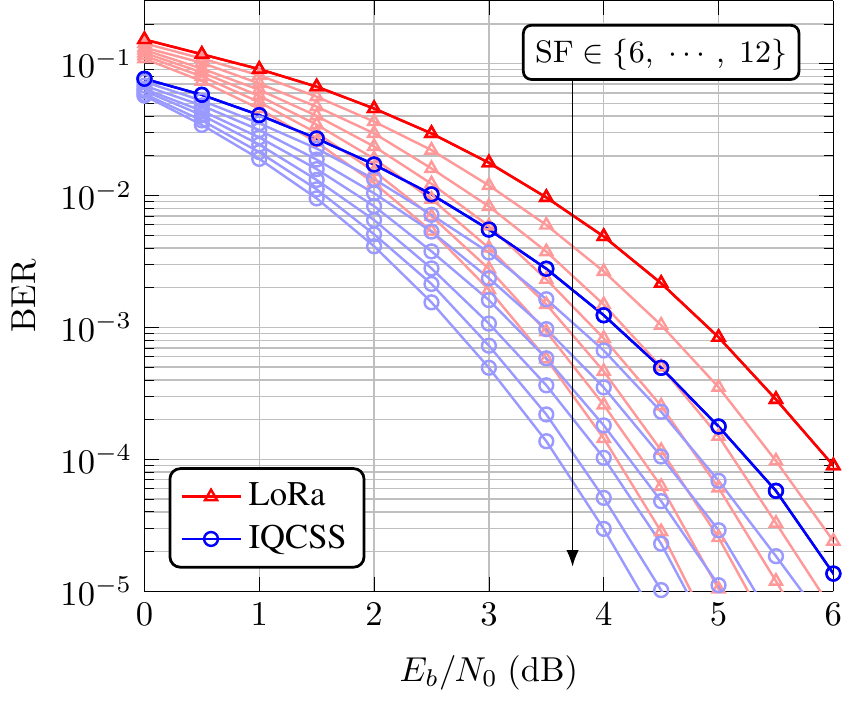}
		\caption{Bit energy versus bit error ratio under AWGN channel.}
		\label{awgn}
	\end{figure}
	
	\begin{figure}[ht!]
		\centering
		\includegraphics[]{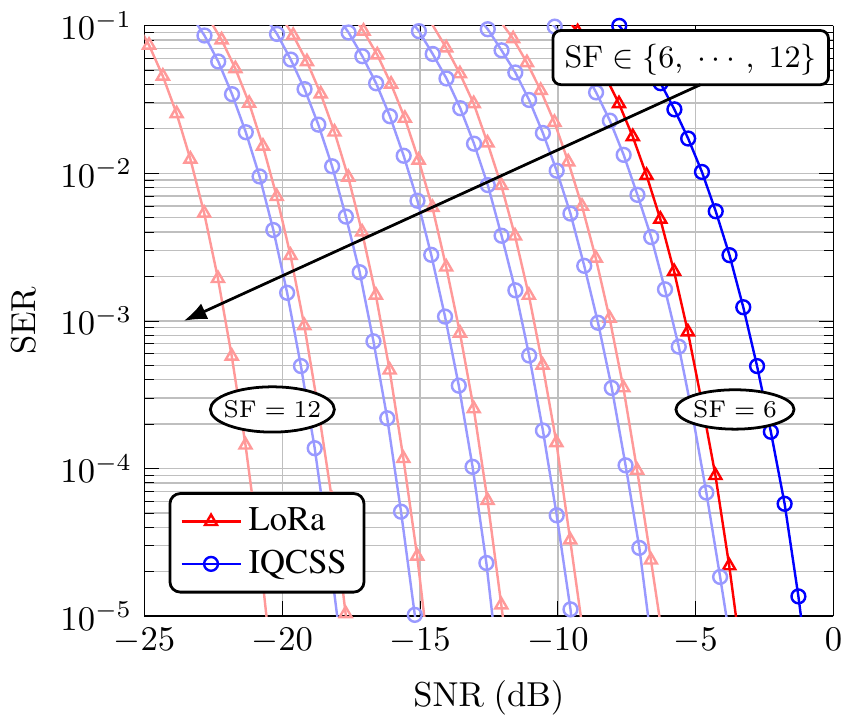}
		\caption{Signal-to-noise ratio versus bit error ratio under AWGN channel.}
		\label{awgnSNR}
	\end{figure}
	
	Figure \ref{awgn} shows the estimated BER under AWGN channel for LoRa and IQCSS. 
	Note that IQCSS actually transmits twice the amount of bits when compared with LoRa.  
	For both LoRa and IQCSS, larger SF will result in better performance, since both use FSK, adding more symbols to the constellation does not reduce the minimum symbol distance.
	There is a gap of about 1 dB between the curves of IQCSS and LoRa for the same SF.
	This is observed because IQCSS collects less noise than LoRa in the process of detection, since it explores the phase information instead of making a decision based solely on the estimated energy of the frequency bins.
	Thus, one can use less energy to transmit more information with the same bit error probability when employing IQCSS over LoRa. 
	For the sake of presentation, darker lines are used for $\mathrm{SF}=6$ in all BER figures.
	
	Figure \ref{awgnSNR} also shows BER under AWGN channel, but taking into consideration the spreading gain.
	The transmit power available is divided between the in-phase and quadrature components of IQCSS, while in LoRa all power is used for a single component.
	Therefore, there is a 3 dB gap between LoRa and IQCSS in Fig. \ref{awgnSNR}. 
	
	\subsection{Throughput analysis}
	The capacity, i.e., maximum achievable throughput under AWGN channel with bandwidth $B$ (Hz) and signal-to-noise ratio ($\mathrm{SNR}$) is given by	$C = B \log_{2}\left( 1 + \mathrm{SNR}\right)$ bits per second (bps). The theoretical bit rate for LoRa is $R_{\mathrm{L}} = \frac{\mathrm{SF}B}{N}$ (bps), and for IQCSS is $R_{\mathrm{IQ}} = \frac{2\mathrm{SF}B}{N}$ (bps). The actual throughput of both schemes for a given $\mathrm{SNR}$ value is related to the symbol error ratio (SER) as $C = (1-\mathrm{SER})R$.
	
	\begin{figure}[ht!]
		\centering
		\includegraphics[]{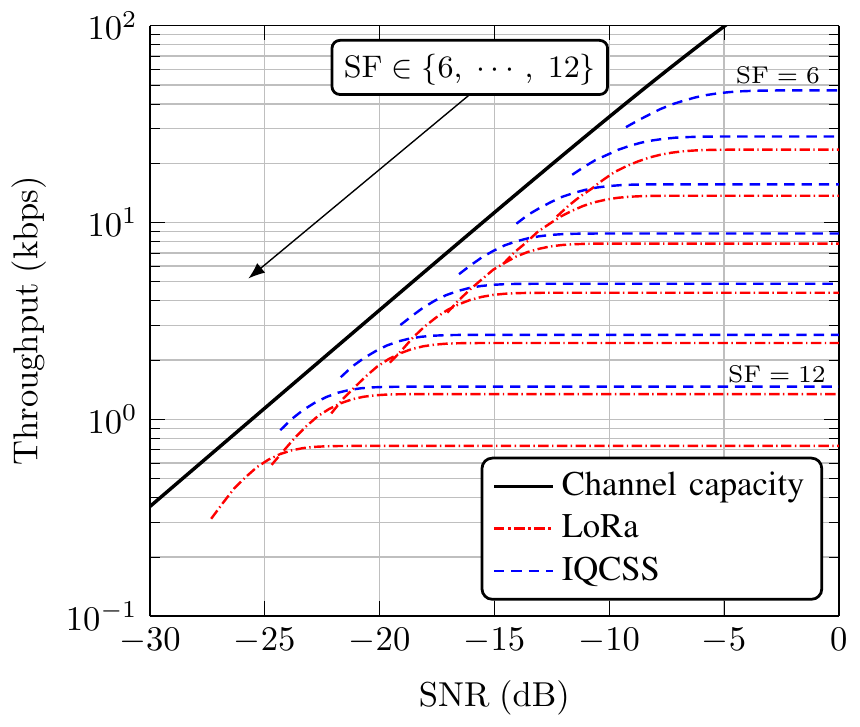}
		\caption{Maximum throughput achievable under AWGN channel.}
		\label{capacity}
	\end{figure}
	
	Figure \ref{capacity} shows the maximum achievable throughput under AWGN channel for LoRa and IQCSS. 
	The Shannon channel capacity is also plotted for reference.
	IQCSS achieves more than double the maximum throughput of LoRa, since it encodes double the amount of bits and benefits from coherent detection at the receiver. 
	Therefore, it makes a more efficient usage of spectrum resources.
	For a given SNR, the throughput is approximately doubled.
	From the perspective of a fixed throughput required by an application, IQCSS can operate in a lower SNR region and, hence, reduce the overall power consumption.
	This are key aspects required for IoT-like applications.
	
	\subsection{Time-variant channel}
	\begin{figure}[t!]
		\centering
		\includegraphics[]{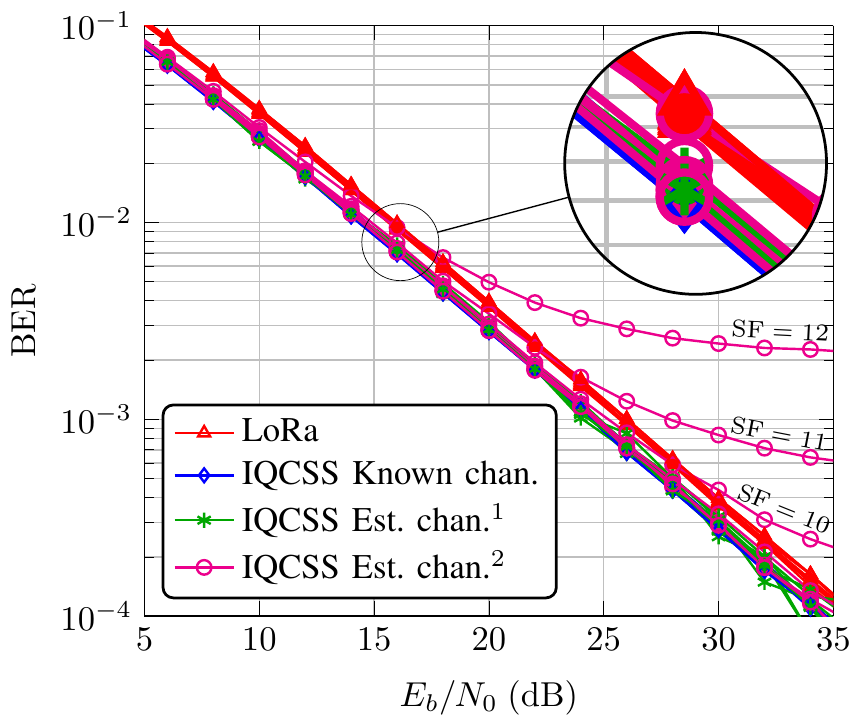}
		\caption{Bit error ratio under Rayleigh channel.}
		\label{rayleigh}
	\end{figure}
	
	Figure \ref{rayleigh} shows the estimated BER under Rayleigh channel for LoRa and IQCSS. 
	The same behavior as observed under AWGN channel regarding energy efficiency is observed between IQCSS and LoRa. 
	Three BER curves are estimated for accessing IQCSS's performance under TV channels. 
	The curve with diamond marks shows the BER considering an unrealistic scenario where the channel coefficient in known at the receiver. 
	For estimating the BER with asterisk marks, we assume that the channel coefficient remains static during the transmission of one frame, which has 8 chirps for synchronization and channel estimation, and 20 chirps encoded with information. 
	These results show that the proposed channel estimation technique yields performance comparable to perfect channel estimation. 
	Lastly, the curve with circle marks considers that the channel coefficient changes during the transmission of one frame. 
	In this last case, we consider that the relative speed between transmitter and receiver is 0.1 km/h, carrier frequency is 863 MHz, the bandwidth occupied is 250 kHz.
	As a result from the mobility, IQCSS suffers a performance degradation when compared with LoRa. 
	However, considering a low mobility scenario, this degradation might be neglected. 
	For this particular simulation scenario, significant performance degradation in observed for SFs greater than 10.
	
	\subsection{Time-variant frequency-selective channel}
	\begin{figure}[t!]
		\centering
		\includegraphics[]{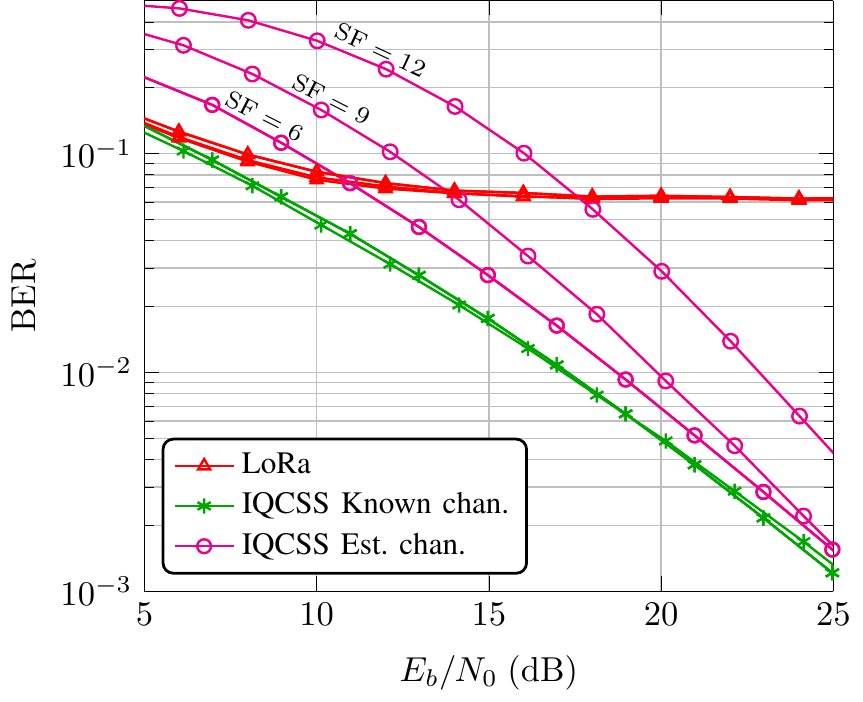}
		\caption{Bit error ratio under TVFS channel.}
		\label{gsmchannel}
	\end{figure}
	
	Figure \ref{gsmchannel} shows the BER under the GSM channel model.
	Due to multipath components of the channel, the signal in Fig. \ref{RE_IM_RX_INDEX} will present multiple peaks, and depending on the channel realization, LoRa's receiver is not able to distinguish between them. 
	Hence, the BER degrades significantly under such scenario.
	Two BER curves are estimated for accessing IQCSS's performance under TVFS channels. 
	The curve with asterisk marks is obtained considering perfect channel estimation, whereas the curve with circle marks considers the LS channel estimation given by (\ref{channel_estimator_selective}).
	For lower $E_b/N_0$ values the LS estimation suffers, since it does not take into account the noise statistics. 
	This will have a negative effect on the BER, specially for larger SF values due to its longer time duration.
	Nevertheless, in the higher $E_b/N_0$ regime, the BER performance approaches the case of ideal channel estimation.
	The mobility scenario is considered for all curves in this subsection.
	Therefore, larger SF values suffer performance degradation, since the channel impulse response changes within the frame duration.
	A CP with length $N_{\mathrm{CP}} = 16$ samples has been employed, together with frequency-domain equalization.
	IQCSS presents acceptable performance particularly for lower SF values.
	
	\section{Conclusion}
	
	This paper presented an alternative modulation scheme inspired by the LoRa PHY for wireless applications that require low energy consumption and larger data throughput. 
	The major advantage of IQCSS lies in the ability to double the SE, and at the same time to improve EE when compared with the conventional CSS employed by LoRa by making use of the phase information contained in the synchronization preamble. 
	With such information, coherent detection at the receiver side becomes possible
	Thus, one observes a performance improvement of about 1 dB in terms of energy efficiency.
	Furthermore, transmission under harsher TVFS channels becomes feasible under the assumption of a CP.
	
	Another interesting aspect of IQCSS is that the receiver is still able to decode the information transmitted using the original LoRa modulation.
	Hence, a gateway designed with IQCSS is backwards compatible with the conventional LoRa.
	
	Future works should explore the performance of IQCSS in combination with multilevel modulation for enhancing even more spectral efficiency.

	\section*{Acknowledgments}
	This research was supported by the European Union under the 5G-RANGE BR-EU project, by DAAD, MESRI, and MEAE under the PROCOPE 2020 project and by the CY Initiative through the ASIA Chair of Excellence.

	\bibliographystyle{ieeetran}
	\bibliography{my_references}
\end{document}